\documentclass[aps,prb,showpacs,superscriptaddress,reprint]{revtex4-1}

\usepackage{graphicx, epsfig, amssymb, amsmath, lineno, subfigure, float,lmodern}

\begin{document}
\pagenumbering{roman}
\title{Tailoring the magnetodynamic properties of nanomagnets using magnetocrystalline and shape anisotropies  }

\author{Vegard Flovik} \email{vegard.flovik@ntnu.no}
\affiliation{Department of Physics, Norwegian University of Science and Technology, N-7491 Trondheim, Norway}

\author{Ferran Maci\`{a}}
\affiliation{Grup de Magnetisme, Dept. de F\'isica Fonamental, Universitat de Barcelona, Spain}

\author{Joan Manel Hern\`{a}ndez}
\affiliation{Grup de Magnetisme, Dept. de F\'isica Fonamental, Universitat de Barcelona, Spain}

\author{Rimantas Bru\v cas }
\affiliation{ Department of Engineering Sciences, Uppsala University, SE-751 21 Uppsala, Sweden}

\author{Maj Hanson}
\affiliation{Department of Applied Physics, Chalmers University of Technology, SE-412 96 Gothenburg, Sweden}

\author{Erik Wahlstr\"om}
\affiliation{Department of Physics, Norwegian University of Science and Technology, N-7491 Trondheim, Norway}

\date{\today}

\begin{abstract}
Magnetodynamical properties of nanomagnets are affected by the demagnetizing fields created by the same nanoelements. In addition, magnetocrystalline anisotropy produces an effective field that also contributes to the spin dynamics. In this article we show how the dimensions of magnetic elements can be used to balance crystalline and shape anisotropies, and that this can be used to tailor the magnetodynamic properties. We study ferromagnetic ellipses patterned from a 10 nm thick epitaxial Fe film with dimensions ranging from $50\times150$  nm to $150\times450$  nm. The study combines ferromagnetic resonance (FMR) spectroscopy with analytical calculations and micromagnetic simulations, and proves that the dynamical properties can be effectively controlled by changing the size of the nanomagnets. We also show how edge defects in the samples influence the magnetization dynamics. Dynamical edge modes localized along the sample edges are strongly influenced by edge defects, and this needs to be taken into account in understanding the full FMR spectrum.

\end{abstract}


\maketitle

\section{Introduction}

The magnetodynamic properties of  nanostructures have received extensive attention, from both fundamental and applications viewpoints \cite{magnetism_review,magnetism_review2,magnetism_review3,magnonics_review,magnonics_review2}. 
Nanometer sized magnetic elements play an important role in advanced magnetic storage schemes \cite{magn_recording,mram}, 
and their static and most importantly their dynamic magnetic properties are being intensely studied. 
While technological applications are important, there is also significant interest in understanding the fundamental behavior of magnetic materials when they are confined to nanoscale dimensions. In confined magnetic elements, there is a complex competition between exchange, dipolar and anisotropic magnetic energies. Understanding the interplay between the various energy terms is thus of importance when investigating the magnetodynamics of such systems.

The magnetization dynamics in patterned magnetic structures has been extensively studied previously \cite{nanomag,nanomag1,nanomag2,nanomag3,nanomag4,nanomag5,nanomag6}. The spin dynamics in elliptical permalloy dots were investigated by Gubbiotti \textit{et al.}\cite{nanomag1}. They studied the various excitation modes as a function of dot eccentricity and in-plane orientation of the applied field, showing how the shape of the ellipses affects the spectrum of excitable modes and their frequencies. 

However, the above mentioned studies of patterned magnetic structures were all performed for systems having a negligible magnetocrystalline anisotropy. Material systems with a significant crystalline anisotropy produce an effective field which also contributes to the spin dynamics. The combination of shape and crystalline anisotropy results in a complex energy landscape, where the interplay of these energy terms determines the magnetodynamic properties of the system.

The influence of shape and crystalline anisotropy on magnetic hysteresis and domain structures in submicron-size Fe particles have previously been investigated by M. Hanson \textit{et al.} \cite{switching}. However, to the best of our knowledge, the dynamic properties of magnetic structures utilizing both crystalline and shape anisotropies remains unexplored. The goal of this study is thus to investigate a system where the energy terms from both crystalline and shape anisotropy contribute to determine the dynamics of the system.

We have investigated a system utilizing epitaxial Fe as the ferromagnetic (FM) material, patterned to an array of elliptical nanomagnets. This results in a system combining the cubic crystalline anisotropy of Fe with the shape anisotropy due to the elliptical shape of the confined magnetic elements.

The dynamic properties were investigated by ferromagnetic resonance (FMR) experiments for ellipses with a thickness of 10 nm and lateral dimensions of $50 \times 150$ nm,  $100 \times 300$ nm and  $150 \times 450$ nm.
The experimental results are compared with micromagnetic simulations, and a macrospin model considering the total free energy density of a ferromagnetic structure containing both crystalline and shape anisotropies. The macrospin model is then used to explore the properties of ellipses with lateral dimensions ranging from $50 \times 150$ nm to $500 \times 1500$ nm, showing how the ellipse size governs the balance between crystalline and shape anisotropy. 

During the fabrication of such structures, the magnetic properties may be affected by edge defects and shape distortions \cite{edge, edgedef,edgedef2,edgedef3}. As the size of the magnetic elements are reduced, the edge regions become increasingly important. Understanding how edge defects affect the magnetodynamic properties of the elements is thus of importance in nanomagnets, where the edge region covers a significant amount of the total sample area. We show how this affects the magnetization dynamics, and that edge defects need to be taken into account in understanding the full FMR spectrum.

\section{Experimental setup}\label{experimental}
The samples are based on a single crystalline Fe film epitaxially grown on MgO(001) substrates. The ferromagnetic ellipses were patterned by e-beam lithography and ion beam milling from a 10 nm thick Fe layer, and have lateral dimensions of  $50 \times 150$ nm, $100 \times 300$ nm and $150 \times 450$ nm. 
The crystalline easy axis [100] and [010] of the Fe film are oriented along the long/short axis of the ellipses, as indicated in Fig. \ref{fig:geom}a. Further details concerning sample growth and processing are similar to that described earlier \cite{sampleprep}.

The FMR experiments were performed using two complementary setups. The cavity FMR measurements were carried out in a commercial X-band electron paramagnetic resonance (EPR) setup with a fixed microwave frequency of 9.4 GHz (Bruker Bio-spin ELEXSYS 500, with a cylindrical TE-011 microwave cavity). The magnitude of the external field is then swept to locate the resonance field, $H_\text{R}$. The sample is attached to a quartz rod connected to a goniometer, allowing to rotate the sample 360 degrees in order to accurately resolve the angular dependence. The FMR measurements were performed with a low amplitude ac modulation of the static field, which allows lock-in detection to be used in order to increase the signal to noise ratio. 

For the broadband FMR measurements, we used a vector network analyzer (VNA) FMR setup with a coplanar waveguide (CPW) excitation structure. The static external field, $H_0$, was applied in the sample plane, and perpendicular to the microwave field from the CPW. This was used to obtain the standard microwave S parameters as a function of frequency for various fixed values of the static field. This allows for a complete field versus frequency map of the resonance absorption, not being limited to a fixed frequency as for the cavity measurements. Data was then collected in a field range of $\pm$ 500 mT, and a frequency range of 1-25 GHz. Typical absorption maps had a step size of $\Delta f = 0.1$ GHz and $\Delta H_0= 5$mT.   

\section{micromagnetic simulations}

The micromagnetic calculations were performed using MuMax \cite{mumax}.
The simulated ellipses have a dimension of $150 \times 450$ nm, with a thickness of 10 nm. In order to have mesh independence, the discretization cells should have sides of the same order, or less than, the two characteristic magnetic length scales of the system. The exchange length, $l_{\text{exch}}=(\frac{A}{K_1})^{1/2}$ and the magnetostatic exchange length $l_{\text{dem}}=(\frac{A}{K_d})^{1/2}$. Here $A$ is the exchange stiffness constant, $K_1$ the first order anisotropy constant, $K_d$ the energy density of the stray field, and an upper limit for $K_d$ is given by $\frac{1}{2}\mu_0 {M_s}^2$. 

Material parameters used in the simulations are standard literature values, with a saturation magnetization, $M_s = 1.7 \times 10^6$ A/m and a crystalline anisotropy constant of $K_1=4.3 \times 10^4 $J/$\text{m}^3$, with the easy axis oriented along the long and short axis of the ellipse. The exchange stiffness was set to a value off A $=21 \times 10^{-12}$ J$\text{m}^{-1}$, and the damping coefficient to $\alpha = 0.01$.

 Performing simulations for a 3d model and a 2d model we obtained the same results, and varying the grid size it was found that the results converge at a grid size of $2 \times 2$ nm. To save computation time the simulation model was thus implemented as a 2d model with a grid size of $2 \times 2$ nm, which is well below the characteristic magnetic length scales of the system ($l_{\text{exch}}= 21$ nm, $l_{\text{dem}}= 3.5$ nm.)

Simulations of the FMR spectrums were performed by using a field relaxation process. The system is first initialized at zero applied field.
If a static field, $\bold{H_0}$, is applied, the simulations are run until the system reaches the new ground state configuration. A 10 mT perturbation field, $\bold{H_p}$ is then applied along the z-axis (out of plane), and the simulation is run until it reaches the ground state configuration for the field $\bold{H_0} + \bold{H_p}$. The  perturbation field is then switched off, allowing the system to relax.
The perturbation causes oscillations of the magnetization around the equilibrium position with a maximum deviation of approx. 1 degrees, avoiding any non-linear effects.  To obtain the resonance frequencies, we take the Fourier power spectrum of the $m_z$ component the first 10 ns of the magnetization relaxation. 
The various excitation modes of the system will then appear as distinct peaks in the Fourier spectrum. \cite{nanomag}

Simulations with an ac field of varying frequency as the perturbing field were also performed, and we obtained the same results as for the field relaxation procedure. The ac approach is however more time consuming, as one has to scan the full frequency range for each value of the applied static field in order to locate all the resonances. To obtain the full field vs. frequency map of the excitation modes in the system we thus used the field relaxation process.

\section{Free energy density and theoretical FMR spectrum}\label{macrospin}

Due to the size and shape of the ellipses, we consider the individual magnetic elements to be in a single domain state. This was also confirmed by MFM imaging of similar samples \cite{dipolar}, where all particles were found to be in a single domain state for a thickness of 10 nm. Increasing the thickness makes it energetically favorable to form flux closure domains, and already at a thickness of 30 nm some of the particles were found to be in such multi domain states. This means that to make sure the magnetic elements are in single domain states, one has to keep the film thickness well below 30 nm for ellipses of the dimensions we have investigated. Having a single domain state allows us to use an analytical macrospin model to investigate the ferromagnetic resonance properties of the system. 

The array of ellipses has an inter-particle spacing of two times the corresponding ellipse dimension in each direction, as illustrated in Fig.\ \ref{fig:geom}a. This spacing is sufficient to significantly reduce the dipolar coupling between the individual elements, and as a first approximation we consider the ellipses as uncoupled magnetic elements. 

We start by defining the geometry of the system, and consider the free energy density of the individual magnetic elements.  From the sample geometry illustrated in Fig.\ \ref{fig:geom}b (magnetic element in the x-y plane), one gets that:  

\begin{figure}[h]
\centering
\includegraphics[height=55mm, width=80mm]{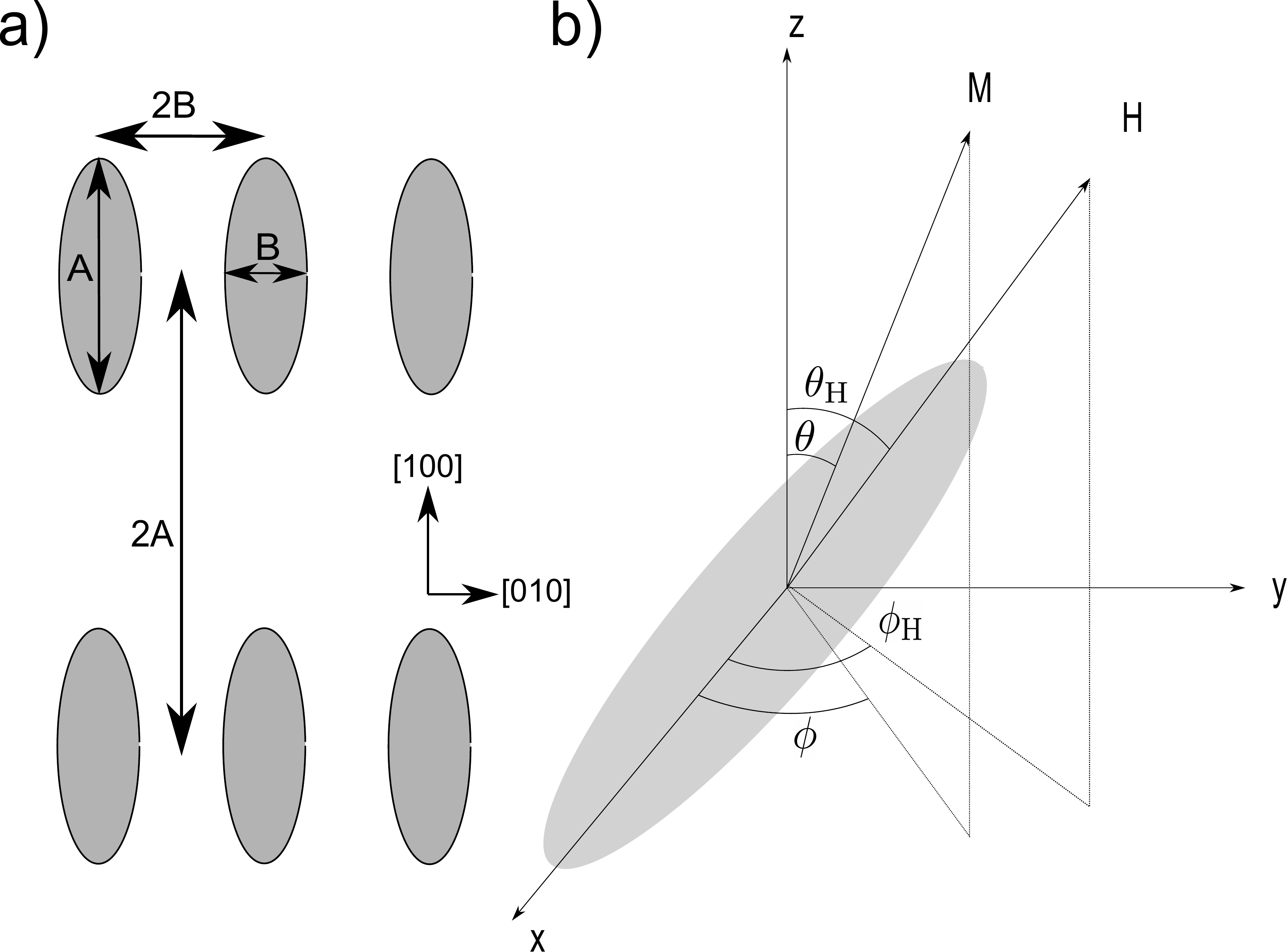} 
\caption{\footnotesize a) Array of ellipses with dimension $A \times B$, an aspect ratio of $A/B=3$ and inter-particle spacing of two times the corresponding ellipse dimension in each direction. The [100] and [010] crystallograpic axis of Fe is oriented along the long/short ellipse axis. b) Field geometry of the individual ellipses}
\label{fig:geom}
\end{figure}

\begin{equation}
\begin{split}
M_x & =M_s \sin \theta \cos \phi \\
M_y & =M_s \sin \theta \sin \phi \\
M_z & =M_s \cos \theta ,
\end{split}
\end{equation}

where $M_s$ is the saturation magnetization. Assuming the external applied field, $\bold{H_0}$, is oriented in the sample plane, $\theta_H = \pi /2$, gives 

\begin{equation}
\begin{split}
H_x & =H_0 \cos \phi_H \\
H_y & =H_0 \sin \phi_H \\
\end{split}
\end{equation}

After defining the geometry, one can calculate the free energy density of the system by adding up the various energy terms. Using a macrospin model, we do not consider the exchange energy. The total free energy density of the system is then given by, $E_{\text{tot}}= E_{\text{Zeeman}} + E_{\text{Demagnetization}} + E_{\text{Anisotropy}}$.  

\begin{equation}
\begin{split}
E_\text{Z} & =-\vec{M} \cdot \vec{H} \\
&= -M_s H_0 [ \sin \theta \cos \phi \cos \phi_H + \sin \theta \sin \phi \sin \phi _H] \\
& = -M_s H_0 \sin \theta \cos(\phi - \phi_H),
\end{split}
\end{equation}

\begin{equation}
\begin{split}
E_\text{Demag} & =\frac{\mu_0}{2} [N_x {M_x}^2 + N_y {M_y}^2 + N_z {M_z}^2] \\
& = \frac{\mu_0 {M_s}^2}{2}  \Big[N_x \sin^2 \theta \cos^2 \phi + N_y  \sin^2 \theta \sin^2 \phi \\
&   +  N_z \cos^2 \theta \Big],
\end{split}
\end{equation}

\noindent
where $\mu_0$ is the vacuum permeability, $N_i$ are the de-magnetization factors and $N_x + N_y + N_z=1 $.  The units for the saturation magnetization and magnetic field are $[M_s]=$A/m and $[H]=$T respectively. 
We assume cubic crystalline anisotropy for the epitaxial Fe film, with the easy axis oriented parallel to the long/short axis of the ellipse, as indicated in Fig. \ref{fig:geom}a. The lowest order term in the crystalline anisotropy energy is then the fourth order term: 

\begin{equation}\label{eq:anis}
\begin{split}
E_\text{Anis} & =K_1 [\alpha_x^2 \alpha_y^2 + \alpha_y^2 \alpha_z^2 + \alpha_z^2 \alpha_x^2] \\
& = K_1 \Big[ \sin^4 \theta \sin^2 \phi \cos^2 \phi  + \sin^2 \theta \sin^2 \phi \cos^2 \theta \\
& +  \sin^2 \theta \cos^2 \phi \cos^2 \theta \Big ],
\end{split}
\end{equation}

where $K_1$ is the magnetocrystalline anisotropy constant and $\alpha_i=M_i/M_s$. After adding the terms, one can write the total free energy density as: 

\begin{equation}\label{eq:energy}
\begin{split}
E_\textrm{tot} &= -M_s H_0 \sin \theta \cos(\phi - \phi_H) \\ 
& + \frac{\mu_0 {M_s}^2}{2}\Bigg [ \sin^2 \theta \cos^2 \phi \left (N_x + \frac{2 K_1}{\mu_0 M_s^2}\sin^2 \theta \sin^2 \phi \right) \\
& +  \sin^2 \theta \sin^2 \phi \left(N_y + \frac{2 K_1}{\mu_0 M_s^2}\cos^2 \theta \right) \\
& + \cos^2 \theta \left(N_z + \frac{2 K_1}{\mu_0 M_s^2}\sin^2 \theta \cos^2 \phi\right) \Bigg ]. 
\end{split}
\end{equation}

Equation (\ref{eq:energy}) describes a complex energy landscape, with competing energies from the various terms.
It is important to note that the orientation of the magnetization, given by $\phi$, might not be parallel to the applied field, $\phi_H$. Thus, to investigate the resonance conditions of the system one must first find the equilibrium orientation of the magnetization. The equilibrium orientation was found by minimizing the free energy density of the system given by Eq.(\ref{eq:energy}) for each value of $H$ and $\phi_H$, and was performed numerically.
After obtaining the equilibrium orientation of the magnetization, one can calculate the resonance frequency $\omega$, given by: \cite{FMR_freq}

\begin{equation}\label{eq:omega}
\omega= \frac{\gamma}{ \mu_0 M_s \sin \theta} \sqrt{ \left(    \frac{  \partial^2 E_{tot} }{\partial \theta^2  }   \frac{\partial^2 E_{tot} } {\partial \phi^2} - \left( \frac{\partial^2 E_{tot} } {\partial \theta \partial \phi}\right)^2  \right ) }. 
\end{equation}

By solving Eq.(\ref{eq:omega}), one can obtain the resonance frequency as a function of magnitude and direction of the applied field, $\omega(H,\phi_H)$. Calculating the various terms in Eq.(\ref{eq:omega}), one obtains:

\begin{equation} \label{eq:theta3}
\begin{split}
\frac{\partial^2 E_{tot}} {\partial \theta^2} &  =  M_s H  \sin \theta \cos(\phi-\phi_H) \\
& +\frac{K_1}{4} \Bigg[  \cos 2\theta \Big( 1-\cos 4 \phi   - 4 \mu_0 M_s^2 N_z /K_1   \\
&    +  \frac{2 \mu_0 M_s^2}{K_1}( N_x + N_y + (N_x - N_y)\cos 2\phi)\Big) \\
&  + (\cos 4\phi + 7) \cos 4\theta \Bigg],
\end{split}
\end{equation}

\begin{equation} \label{eq:phi3}
\begin{split}
\frac{\partial^2 E_{tot}} {\partial \phi^2}  &=  M_s H \sin \theta \cos(\phi-\phi_H) \\
&+   2K_1 \sin^2 \theta \Bigg[ \cos 4\phi \sin^2 \theta   \\  
&+       \frac{\mu_0 {M_s}^2 (N_y-N_x)}{2K_1} \cos 2\phi   \Bigg],
\end{split}
\end{equation}

\begin{equation} \label{eq:thetaphi2}
\begin{split}
\frac{\partial^2 E_{tot}} {\partial \theta \partial \phi} & =   M_s H \cos \theta \sin(\phi-\phi_H)  \\
&+  8K_1 \sin \phi \cos \phi \sin \theta \cos \theta \Bigg[ \cos 2\phi \sin^2 \theta \\
&+ \frac{  \mu_0 M_s^2   (N_y - N_x)}{4K_1}    \Bigg].
\end{split}
\end{equation}

For thin films, one can simplify these expressions by assuming that the magnetization is oriented in the film plane, $\theta = \pi/2$. After introducing the anisotropy field, $H_k =2 K_1/ M_s$, one obtains the resonance frequency given by Eq.(\ref{eq:omega}):

\begin{equation}\label{eq:res4}
\begin{split}
\left(\frac{\omega}{\gamma}\right)^2 & = \bigg[ H  \cos(\phi-\phi_H) + \mu_0 M_s \Big(N_z \\
&- (\frac{N_x + N_y + (N_x - N_y)\cos 2\phi}{2})\Big) + \frac{H_k}{4}(3 + \cos 4\phi)\bigg] \\
& \times \bigg[ H \cos(\phi-\phi_H) + H_k \cos 4 \phi \\
&+ \mu_0 M_s (N_y - N_x) \cos 2\phi \bigg].
\end{split}
\end{equation}

Equation (\ref{eq:res4}) gives the resonance frequency for the general case, with the assumption that the magnetization is oriented in the sample plane. Depending on the shape and size of the magnetic elements, one can then adjust the demagnetization factors $N_i$ to obtain the resonance conditions for various samples.

In addition to the four-fold symmetry from the cubic anisotropy, one notices that in this case there are additional terms of two-fold symmetry due to the shape anisotropy along the long/short axis of the ellipse. The resonance conditions of the system are thus more complicated, and are determined by the interplay of shape and crystalline anisotropies. This brings us to the main topic of the study, to investigate how tuning the various energy terms changes the magnetodynamic properties of the system.

\section{Results and discussion}

\subsection{Cavity FMR measurements}

\begin{figure}[h]
\centering
\includegraphics[height=50mm, width=90mm]{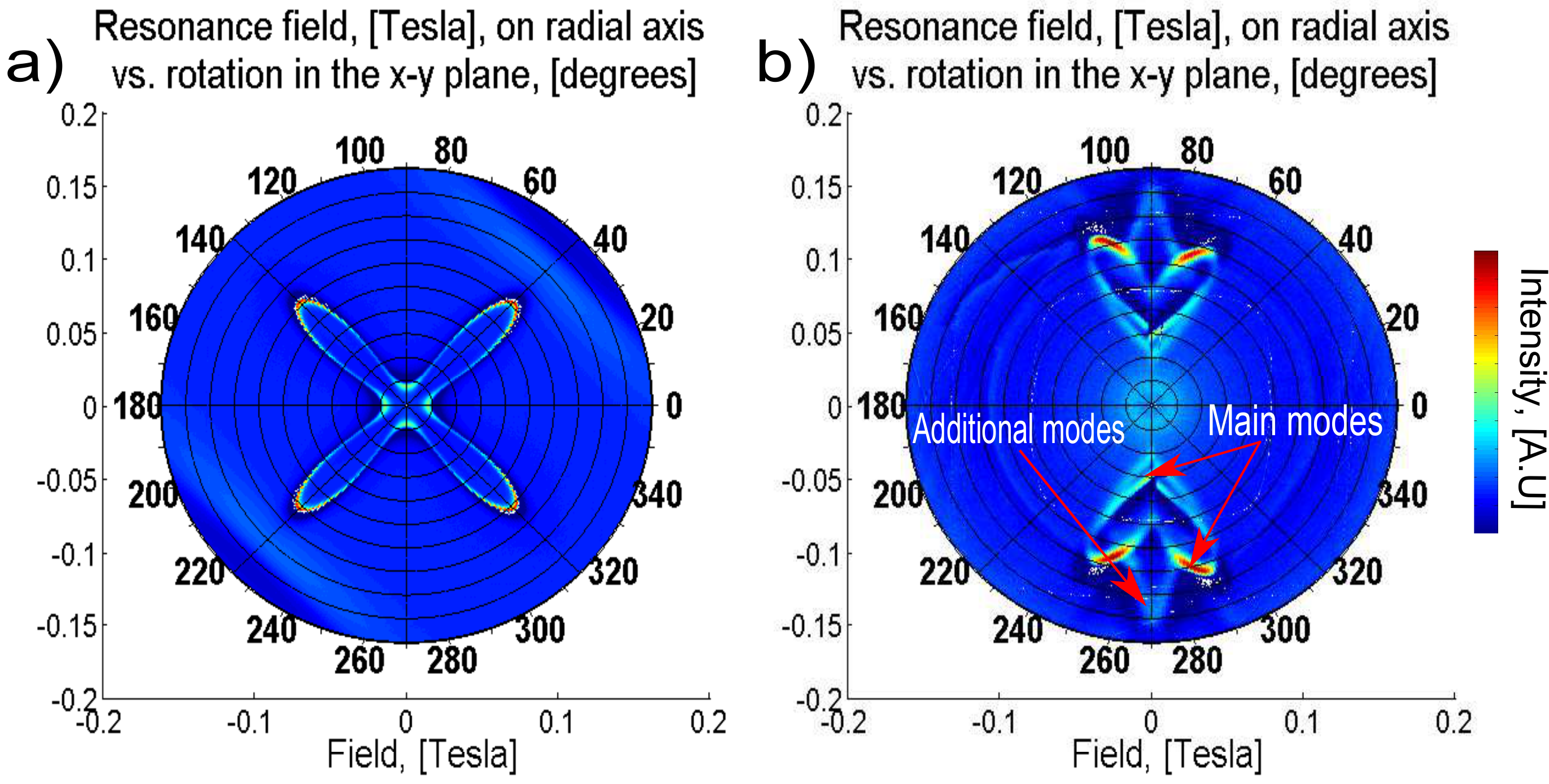} 
\caption{\footnotesize (Color online) Experimental FMR spectrum for a) continuous film and b) ellipses of dimension $150 \times 450$ nm from the X-band cavity FMR setup. }
\label{fig:exp}
\end{figure}

The experiments to investigate the angular dependence were performed in the X-band cavity FMR setup described in section \ref{experimental}. This gives an angular FMR spectrum for both the continuous film and an array of ellipses of dimension $150 \times 450$ nm, as shown in  Fig.\ \ref{fig:exp}.  

Going from a continuous film to a patterned array of ellipses, there is a significant difference. For the continuous film, the four-fold symmetry due to the cubic crystalline anisotropy in Fe is dominating.  
For the ellipses the situation is more complicated, as there are competing energies also from the shape anisotropy. 

\begin{figure}[h]
\centering
\includegraphics[height=90mm, width=80mm]{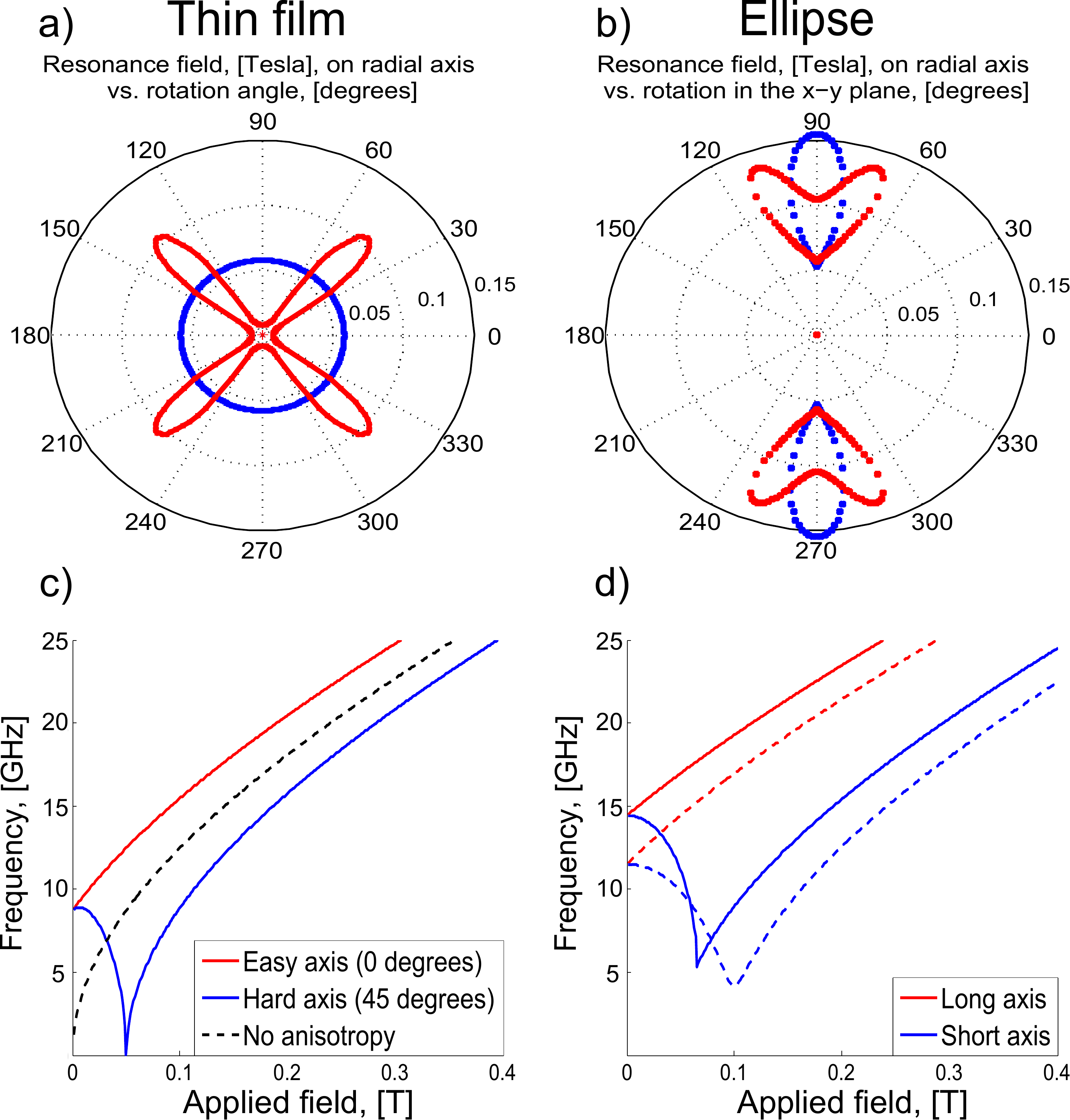} 
\caption{\footnotesize (Color online) Upper figures: Theoretical data for resonance field versus rotation angle for a) continuous film and b) ellipse of dimension $10 \times 150 \times 450$ nm, with (red) and without (blue) crystalline anisotropy. Lower figures: Dispersion for c) continuous film and d) ellipse of dimension $10 \times 150 \times 450$ nm, with (solid lines) and without (dotted lines) crystalline anisotropy.}
\label{fig:theory}
\end{figure}

To investigate this, we compare the experimental and theoretical results. By solving Eq.(\ref{eq:res4}) after first minimizing the free energy density for each value of $H$ and $\phi_H$, one gets the FMR dispersion relations shown in the lower panel of Fig.\ \ref{fig:theory}. From Eq.(\ref{eq:res4}), the relevant parameters determining the dispersion are the demagnetization factors $N_i$, the anisotropy field $H_k$ and the saturation magnetization $M_s$. 
In nanometer-dimension magnetic structures, estimates of the demagnetization factors using an ellipsoidal formulae are considered to represent the anisotropy fields well \cite{demag_factors, demag_factors2}. 
The factors $N_i$ were found from \cite{demag_factors}, and for an ellipse of dimension $10 \times 150 \times 450$ nm they are: $N_x \approx 0.005$, $N_y\approx 0.05$ and $N_z=1-N_x - N_y$. The anisotropy field $H_k$ was determined from the experimental FMR spectrum in Fig.\ \ref{fig:exp}a, and was found to be approx. 50 mT. 
In the calculations, $M_s$ was adjusted to obtain the best fit between the experimental and theoretical spectrum, and the best fit was found for $M_s=1.5 \times  10^{6}$ A/m  (a reduction of approx. 10\% compared to textbook values of $M_s$ for Fe).

To compare the angular dependence of the theoretical spectrum with experimental results from the cavity measurements shown in Fig.\ \ref{fig:exp}, one can inverte the solution. This rather gives the resonance field $H_R$, as a function of rotation angle for a fixed excitation frequency of 9.4 GHz, and the inverted solution is shown in the upper panel of Fig.\ \ref{fig:theory}. To distinguish the effect of crystalline anisotropy and shape anisotropy, the same calculations were also performed assuming polycrystalline Fe, setting $H_\text{k}=0$.

Comparing theory and experiment in Fig.\ \ref{fig:exp} and \ref{fig:theory}, one notices that for the continuous film, both show the expected four-fold cubic symmetry. For the ellipses, the theory replicates the "heart shape" of the resonance well. In the experimental data in Fig.\ \ref{fig:exp}b, there are also some additional weak resonance lines. It is known that regions along the sample edges could lead to a spectrum of additional edge modes \cite{edge,nanomag1,nanomag2}. However, from our experiments we observe that the main mode is dominating, and thus focus on this in the following. The other resonances are characterized and discussed in detail in section \ref{simulations}.

\subsection{Size of the ellipses}
To investigate the interplay of shape anisotropy and crystalline anisotropy, we studied ellipses of various lateral dimensions but with the same aspect ratio of 1:3.
Changing the sample size affects the balance between crystalline and shape anisotropy in the free energy density. As shown using our macrospin model for the main FMR mode, this will in turn change the resonance frequency. 
There are two limiting cases worth noticing: in the limit of a very large ellipse, one should expect a behavior close to that of a continuous film, where crystalline anisotropy is dominating. By gradually reducing the size of the ellipse, shape anisotropy becomes increasingly important. This means that one can use the size of the magnetic elements to tune the ratio between crystalline and shape anisotropies, and thus change the magnetodynamic properties of the system. 

\begin{figure}[h]
\centering
\includegraphics[height=80mm, width=90mm]{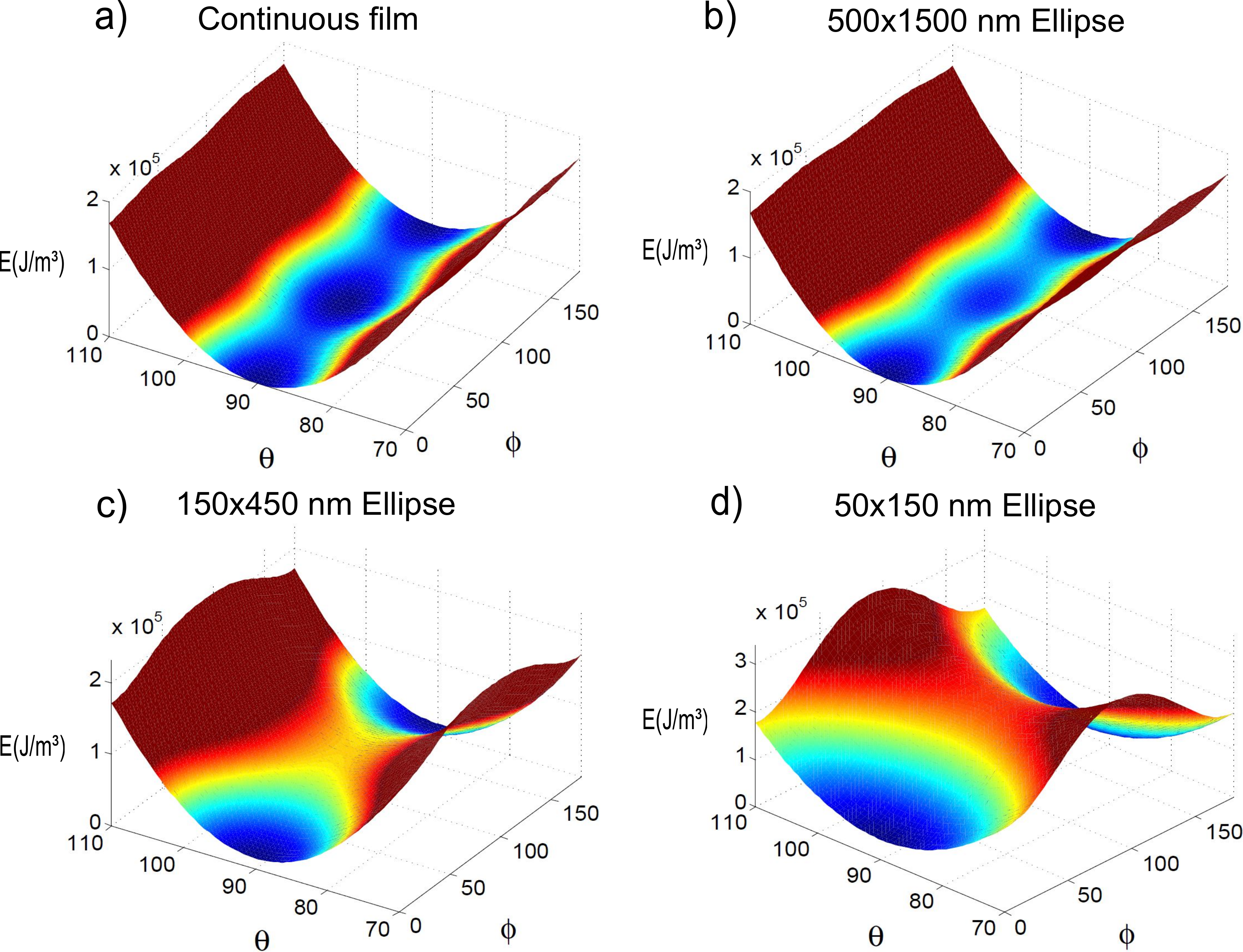} 
\caption{\footnotesize (Color online)  Free energy density given by Eq.(\ref{eq:energy}) for a) continuous film, b) $500 \times 1500$ nm ellipse, c) $150 \times 450$ nm ellipse, d) $50 \times 150$ nm ellipse. Film thickness is 10 nm in all cases. }
\label{fig:size_E_landscape}
\end{figure}

Changing the dimensions of the ellipse affects the free energy density of the system, given by Eq.(\ref{eq:energy}). The transition from a continuous film to a small ellipse can be observed by considering the energy landscape of the system as a function of the ellipse dimension, as shown in Fig.\ \ref{fig:size_E_landscape}. 

Figure \ref{fig:size_E_landscape} indicates  how the free energy density changes when one gradually reduces the size of the ellipse from the upper limit of a continuous film, to an ellipse of dimension $50 \times 150$ nm. As expected, one notices that in all cases the magnetization favors an orientation in the sample plane ($\theta=90$, from sample geometry as defined in Fig.\ \ref{fig:geom}b). For the continuous film and the largest ellipse in Fig.\ \ref{fig:size_E_landscape}a and b, one can clearly see the dominating crystalline anisotropy, with a four-fold symmetry between the energy minima along the $\phi$ axis. 

In the intermediate case for an ellipse of dimension $150 \times 450$ nm, one has two dominating energy minima at $\phi=0$ and $\phi=180$ (magnetization along the long axis of the ellipse). In addition, there is a quite flat saddle point at $\phi=90$ (which corresponds to a magnetization along the short axis of the ellipse). This is not a stable energy minimum, but the flatness of the saddle point means that applying a small magnetic field along this axis will create a local energy minimum along this direction.

For the smallest ellipse, the energy landscape is dominated by the two-fold shape anisotropy along the long axis of the ellipse. To align the magnetization along the short axis of the ellipse ($\phi=90$) will thus require a quite large external field. 

\begin{figure}[h]
\centering
\includegraphics[height=90mm, width=90mm]{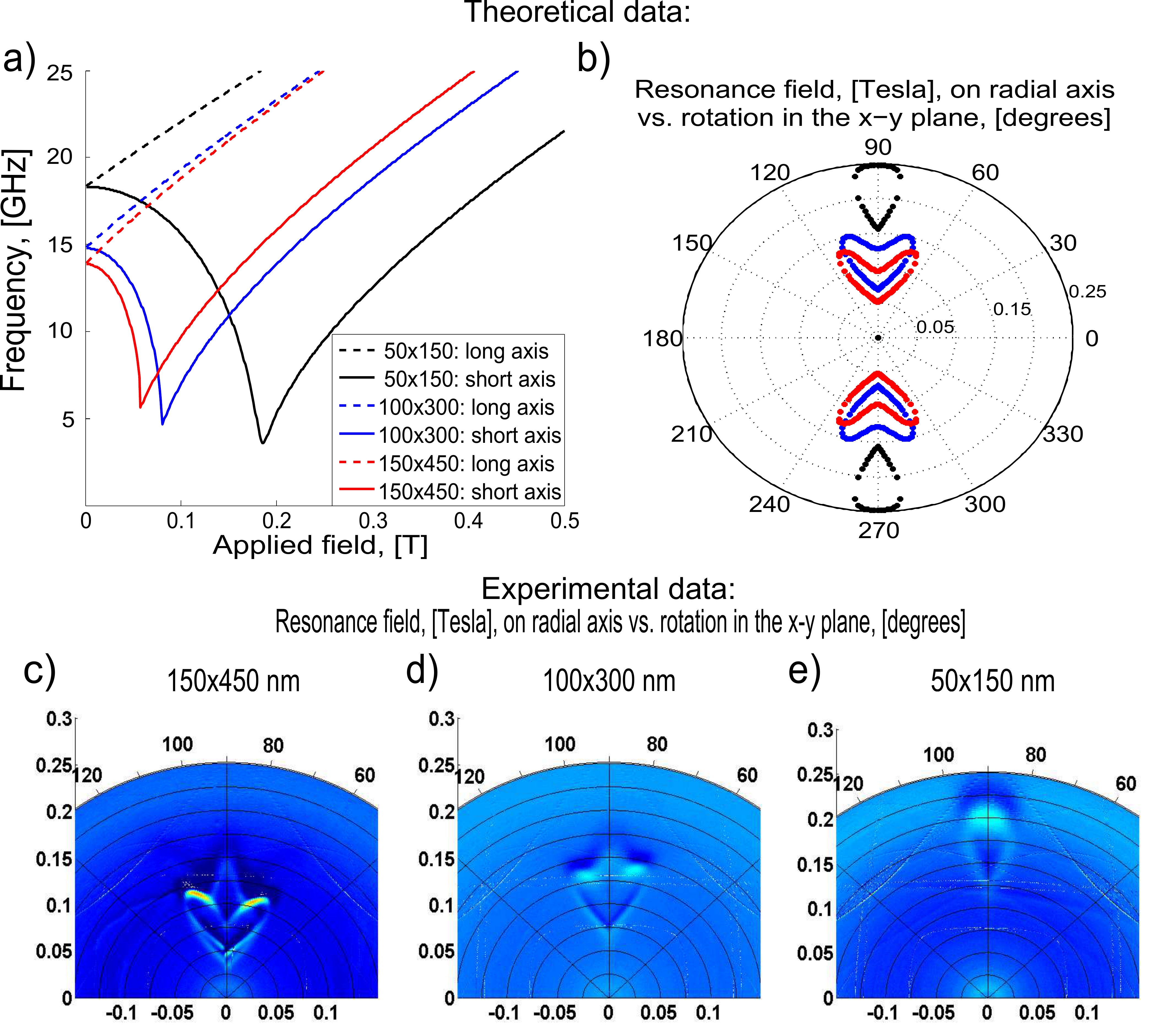} 
\caption{\footnotesize (Color online) a) Theoretical dispersion for ellipses of dimension $50 \times 150$ nm (Black), $100 \times 300$ nm (Blue) and $150 \times 450$ nm (Red). b) Angular dependence of same data. c)  Experimental data for ellipse of dimension $150 \times 450$ nm d) $100 \times 300$ nm and e) $50 \times 150$ nm. }
\label{fig:size_effect}
\end{figure}

As shown in section \ref{macrospin}, the FMR frequency given by Eq.(\ref{eq:res4}) is determined by the free energy density of the system. Adjusting the lateral dimensions of the ellipse is thus an important parameter controlling the FMR frequency. From Eq.(\ref{eq:res4}), one notices that the resonance frequency is determined by contributions of both two-fold and four-fold symmetry. From this expression, the relevant ratio to determine which term will dominate is given by $H_\text{K}/\mu_0 M_s(N_\text{x}-N_\text{y})$. Changing the ellipse dimensions, and thus the demagnetization factors $N_i$, affects the resonance frequency significantly, as shown in the upper panel of  Fig.\ \ref{fig:size_effect}.

Figure\  \ref{fig:size_effect}a and b compare the theoretical FMR spectrum for ellipses of dimension $150 \times 450$ nm, $100 \times 300$ nm and $50 \times 150$ nm. As the dimensions of the ellipse are reduced, the two-fold shape anisotropy tends to dominate over the crystalline anisotropy, and the "heart shape" of the spectrum in Fig.\ \ref{fig:size_effect}b due to the cubic crystalline anisotropy is suppressed. Comparing the theoretical results with the experimental data in the lower panel of Fig.\ \ref{fig:size_effect}, they follow the same trend. As the size is reduced, the resonance is shifted to slightly higher fields, and the "heart shape" of the resonance gets suppressed. 

\begin{figure}[h]
\centering
\includegraphics[height=45mm, width=85mm]{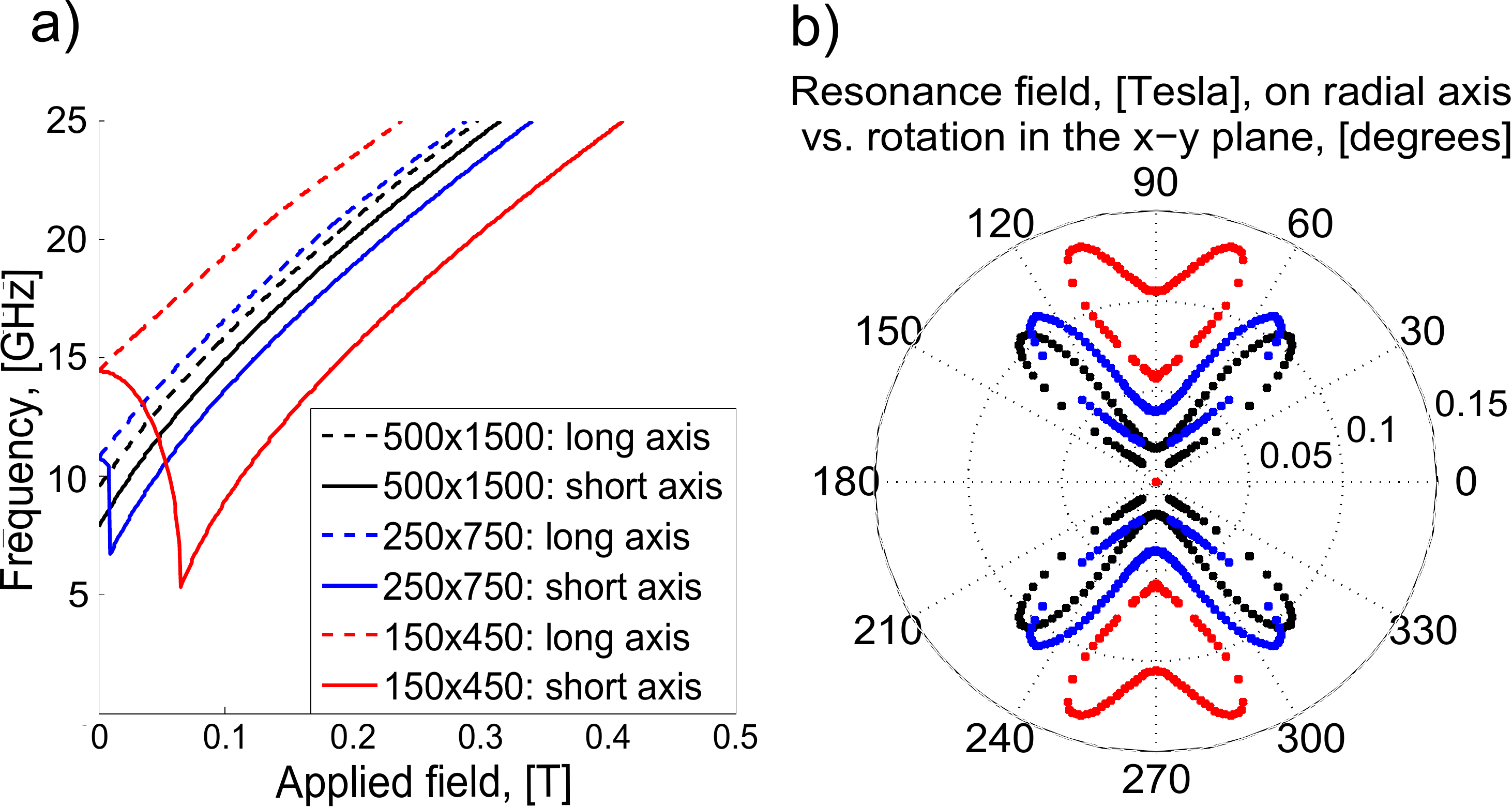} 
\caption{\footnotesize (Color online) a) Theoretical dispersion for ellipses of dimension $500 \times 1500$ nm (Black),  $250 \times 750$ nm (Blue) and $150 \times 450$ nm (Red). b) Angular dependence of same data.  }
\label{fig:size_effect2}
\end{figure}

Investigating the opposite limit, one can determine when the crystalline anisotropy starts to dominate. Comparing the theoretical FMR spectrum for ellipses of dimension $150 \times 450$ nm, $250 \times 750$ nm and $500 \times 1500$ nm in Fig.\ \ref{fig:size_effect2}, one notices that by increasing the size, the effect of shape anisotropy is suppressed compared to that of crystalline anisotropy.

For an ellipse of dimension $500 \times 1500$ nm the dispersion starts to look similar along the long/short axis of the ellipse, as indicated in Fig.\ \ref{fig:size_effect2}a.
If the only contribution was from the crystalline anisotropy, the dispersion should be identical along the long/short axis due to the four-fold symmetry. 
Comparing the FMR spectrum for the largest ellipse in Fig.\ \ref{fig:size_effect2}b to that of a continuous film in Fig.\ \ref{fig:theory}a, they look very similar. This indicates that as the sample dimensions approach the micrometric scale, shape anisotropies play a minor role compared to the crystalline anisotropy. 

To summarize the size dependence, we have shown that for sample dimensions above approx. 1 $\mu m$, crystalline anisotropy will dominate. In the opposite size limit, shape anisotropy will dominate for sample dimensions below approx. $50 \times 150$ nm. In this intermediate regime, one can thus effectively use the sample size as a parameter to tune the balance between crystalline and shape anisotropies.

\subsection{Broadband FMR measurements and micromagnetic simulations}\label{simulations}
The assumption that the magnetization in the individual ellipses is uniform is a good approximation at the center of the ellipse, but along the edges the magnetization will be less uniform due to the demagnetizing fields.
Regions along the sample edges could lead to a spectrum of additional edge modes \cite{edge,nanomag1}. In addition, there could be other spin-wave excitations with non zero wave vectors, and correspondingly varying frequencies \cite{nanomag1,nanomag2}. 
To characterize the various resonances, we thus performed a series of broadband FMR measurement in combination with micromagnetic simulations. 

\begin{figure}[h]
\centering
\includegraphics[height=90mm, width=80mm]{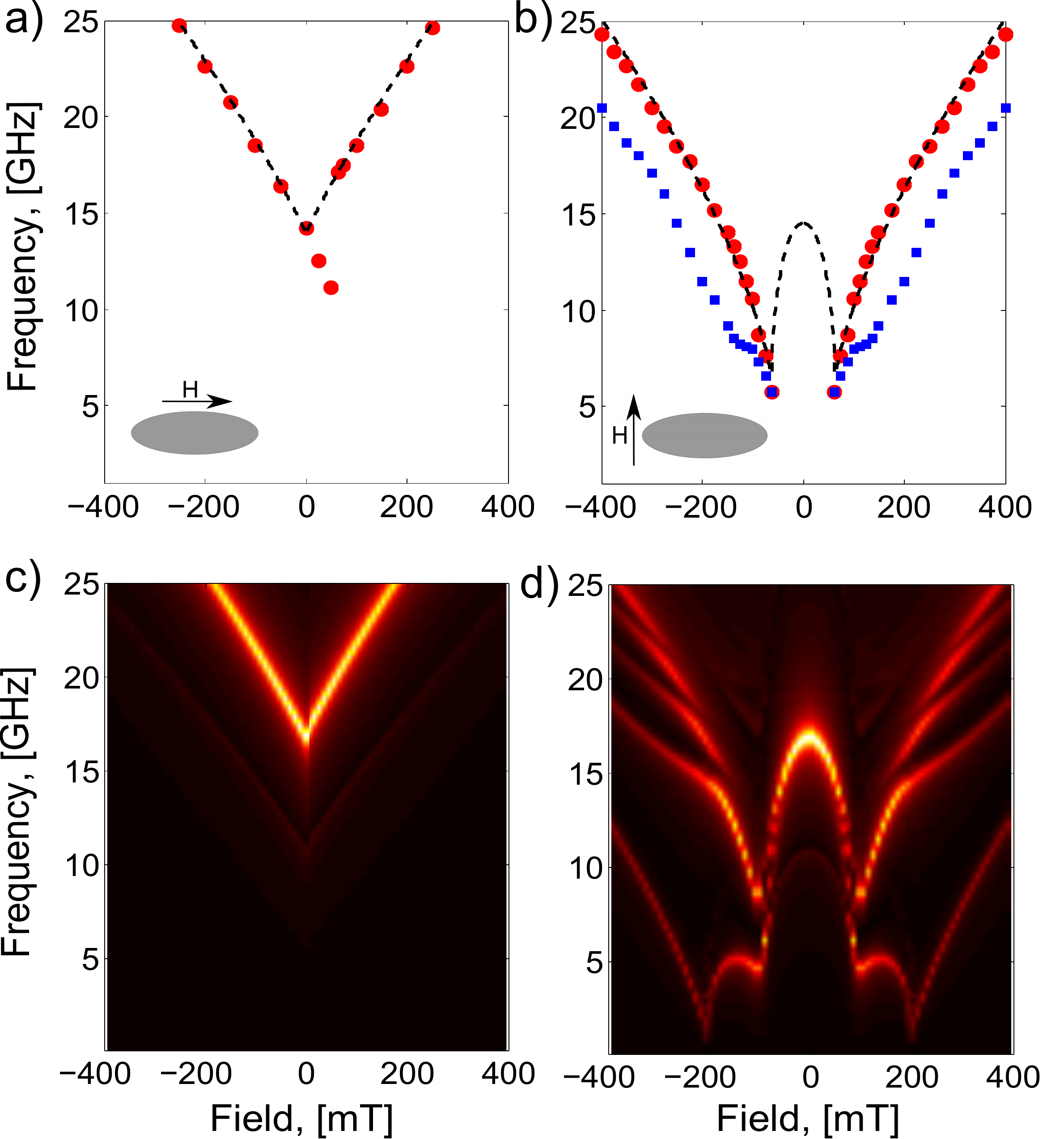} 
\caption{\footnotesize (Color online) Upper panel: Experimental FMR spectrum for ellipses with field oriented along long/short axis. Experimental data is shown as red dots/blue squares, and theoretical spectrum as dotted black line. a) Field sweep from negative to positive field along long axis, also showing the switching of the magnetization at approx. 75mT. b) Similar measurement along the short axis, showing the main mode (red dots) and an additional weak resonance at lower frequency (blue squares). Lower panel: Simulated FMR spectrum for a single ellipse of dimension $150 \times 450$ nm with the field oriented along the c) long and d) short axis.}
\label{fig:theory_vs_exp}
\end{figure}

To obtain a complete field versus frequency map of the FMR absorption, we performed experiments using the broadband setup described in section \ref{experimental}. The experimental FMR absorption peaks were extracted, and are shown in the upper panel of Fig.\ \ref{fig:theory_vs_exp}. Red dots represent the main FMR mode, and the blue squares the additional weaker mode. For clarity only a few selected data points are included, where the uncertainty in determining the absorption peak position is of the order of the dot size. The experimental results are then compared with the theoretical FMR spectrum from the macrospin model, shown as dotted black lines. 

The agreement between theory and experiment is good for an applied field oriented along the long axis of the ellipse, as indicated in Fig.\ \ref{fig:theory_vs_exp}a. Sweeping the field from negative to positive, one also notices the switching of the magnetization. As the field is swept from negative to zero, the FMR frequency decreases as expected. This continues also for positive fields until the external field is strong enough to overcome the anisotropy favoring the magnetization along the long axis of the ellipse. The switching is then observed as an abrubt jump in the FMR spectrum.

When applying the field along the short axis there are two parallel dispersing lines, as shown in Fig.\ \ref{fig:theory_vs_exp}b. A high frequency resonance and an additional weaker resonance at lower frequency, which corresponds well to the additional resonance also seen in the cavity measurements (see  Fig.\ \ref{fig:exp}b).
Comparing the measurements along the short axis with the theoretical dispersion, one does not observe the low field resonance in Fig.\ \ref{fig:theory_vs_exp}b (the black dotted line below 100 mT). 
In this field range the magnetization is not saturated and it is still oriented along the long axis of the ellipse, being parallel to the microwave (MW) pumping field  from the CPW. However, in the cavity measurements we observed both resonances, because the pumping field is, in this case, oriented out of the sample plane and thus  perpendicular to the magnetization. The first resonance is observed at a field of $\sim$ 50 mT (see Fig. \ref{fig:exp}b), and a second one at $\sim$ 100 mT, which agrees well with the expected resonance fields from the theoretical curves shown in Fig. \ref{fig:theory_vs_exp}b at a frequency of 9.4 GHz. At higher fields the magnetization in the ellipse saturates in the direction of the external field, being perpendicular to the MW pumping field from the CPW, and thus the theoretical spectrum corresponds well with the high-frequency branch of the experimental data.

Using a macrospin model, one accounts only for the main FMR mode. In order to investigate the observed low frequency resonance we performed micromagnetic simulations. The model was implemented as a single ellipse of dimensions $150 \times 450$ nm with a thickness of 10 nm, and the simulated FMR spectrums are shown in the lower panel of Fig. \ref{fig:theory_vs_exp}. Comparing the experimental data with the micromagnetic simulations, we notice a few differences. Applying the field along the long axis of the ellipse, the simulated and experimental data both show a single dispersing resonance. The simulated FMR frequency is however noticeably higher than the experimental results. 
Applying the field along the short axis of the ellipse, the differences between the experimental and simulated FMR spectrum are more significant. The experimental data show two parallel dispersing lines, whereas the simulated spectrum shows a whole range of various excitation modes.

A similar splitting of the main mode has been observed experimentally in elliptical permalloy dots, and was attributed to a hybridization of the main mode with other spin-wave modes \cite{nanomag1,nanomag2}. 
A study of the excitation modes in permalloy dots as a function of dot eccentricity has been performed by Gubiotti \textit{et al}. \cite{nanomag1}, where they found a large range of possible modes depending on the orientation of the external field with respect to the axis of the dots. The number of modes in our system compared to theirs may be smaller because of the different material parameters and sample size. The exchange stiffness in Fe is almost twice that of permalloy, and combined with a smaller sample size this results in a reduction in the number of excitation modes due to the increased exchange energy. This was also confirmed in our simulations, where
the mode splitting disappear when reducing the sample size or increasing the exchange stiffness.

\begin{figure}[h]
\centering
\includegraphics[height=40mm, width=85mm]{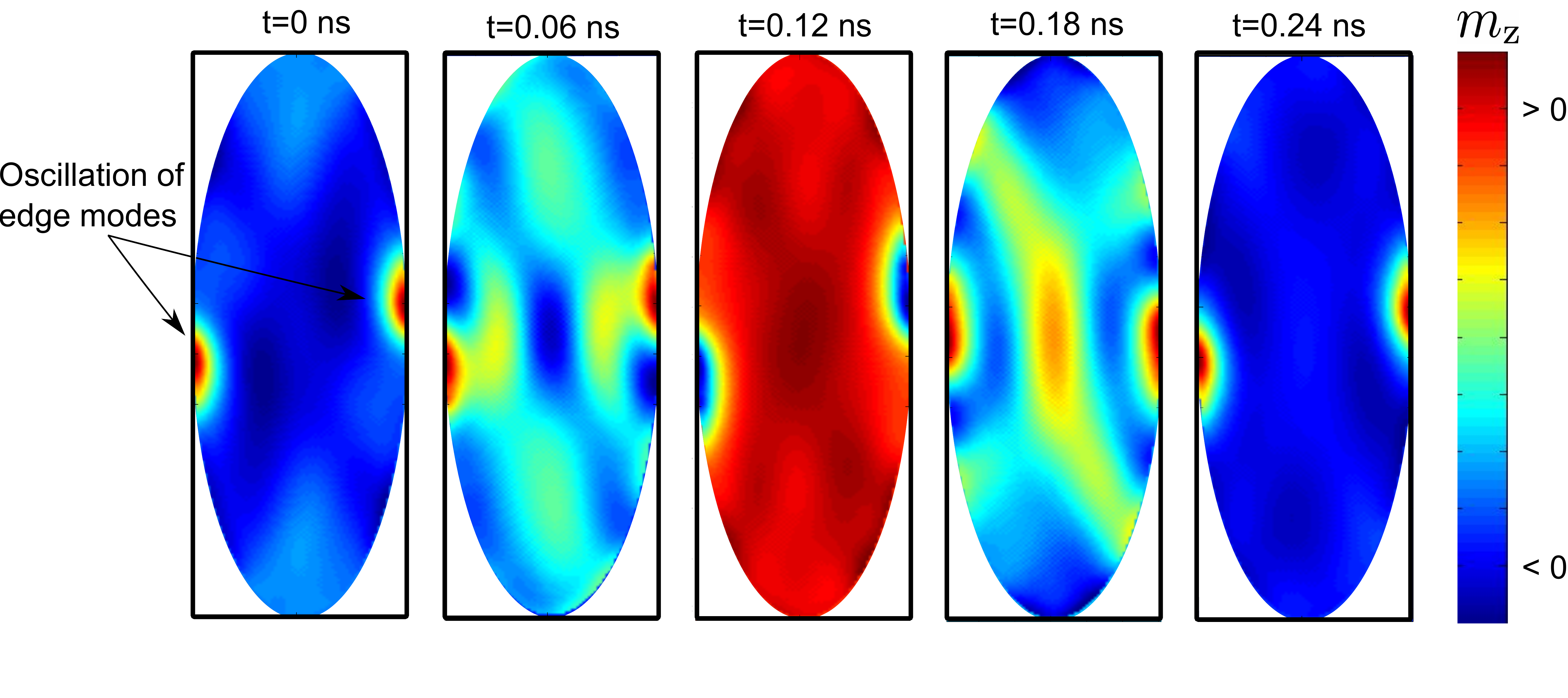} 
\caption{\footnotesize (Color online) $m_z$ component, showing one oscillation period of the edge mode at $H$= 150 mT, corresponding to a frequency of approx. 4 GHz. }
\label{fig:edge_osc}
\end{figure}

The low frequency branch in Fig. \ref{fig:theory_vs_exp}d was identified by imaging the $m_z$ component from the micromagnetic simulations (out of plane component). From the periodic oscillations of the magnetization, we determined the low frequency resonance to be localized along the edges of the ellipse, as indicated in Fig. \ref{fig:edge_osc} for an applied field of 150 mT along the short axis.

After identifying the excitation modes, one needs to consider why there is a significant difference between the simulated and experimental FMR spectrum. It is known that the fabrication process of nanostructures can lead to distortions and defects at the sample edges  \cite{edge, edgedef,edgedef2,edgedef3}. To investigate how this would affect the magnetodynamic properties, the effects of edge defects need to be taken into account in the simulation model.

\subsubsection{ Edge modes and edge defects}

In the initial simulations, the edges of the ellipse were treated as ideal. However, the samples most likely have some kind of non-ideal edges which could influence the FMR spectrum. The effects of non-ideal edges on the dynamics have been investigated theoretically by McMichael \textit{et al.} \cite{edge}. It was shown that several cases such as edge geometry, reduced edge magnetization and surface anisotropy on the edge surface all had similar effects. The main effect was to reduce the edge saturation field, which is the field needed to align the magnetization at the edge nearly parallel to the applied field.
A reduced edge magnetization will also lead to a smaller effective demagnetization field along the edges. This would cause a significant increase of the edge mode resonance frequency compared to that of an ideal edge, and the shift could be in the order of several GHz \cite{edge}. Such effects would be less important when the field is oriented along the easy axis of the ellipse, explaining the better agreement between the simulated and experimental spectrum in that geometry.

\begin{figure}[h]
\centering
\includegraphics[height=100mm, width=75mm]{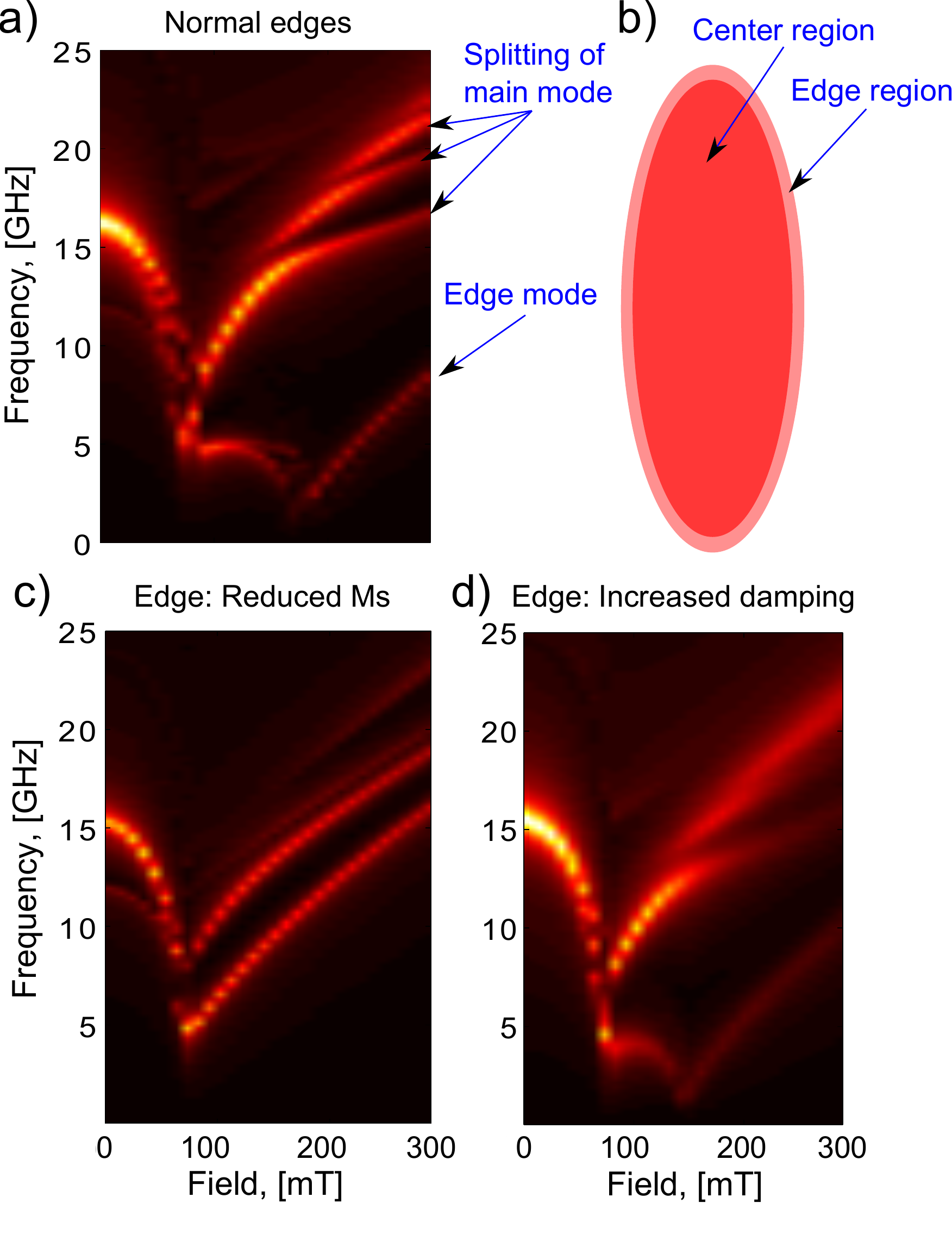} 
\caption{\footnotesize (Color online) a) Simulation using normal edges, showing the edge mode and splitting of the main mode. b) Schematic of simulation model with a defined edge region. c) Simulation with reduced $M_s$ in edge region (reduction of 40 $\%$). d) Simulation with increased damping $\alpha$ in edge region (from 0.01 to 0.1). }
\label{fig:edgedefects}
\end{figure}

To account for edge defects in the simulations, we made a model where the material properties were changed along the edges of the ellipse. In a real sample the variation of the material properties when approaching the sample edge should be gradual,  but as a first approximation the model was defined with two distinct regions. The width of the edge region was set to 10 nm, and is within the same width range as that investigated theoretically by McMichael \textit{et al.} \cite{edge}. A schematic of the model including edge defects is shown in Fig. \ref{fig:edgedefects}b. 

As mentioned in section \ref{experimental}, the samples were defined by ion beam milling. This can affect the magnetic properties of the sample \cite{edgedef3}, and a more disordered edge region could lead to an increased damping of the FMR modes. 
Two kinds of defects have thus been considered in the simulations; increased damping $\alpha$, and reduced $M_s$.

In the initial simulation model with ideal edges, excited spin waves would be reflected at the edges of the sample. This explains the multiple excitation modes observed in the simulations, due to a hybridization of the main mode with other spin-wave modes \cite{nanomag1,nanomag2} (see Fig. \ref{fig:edgedefects}a). 
As a disordered edge region could lead to increased damping of the FMR modes, we introduced an edge region where the damping was increased from $\alpha=0.01$ to $\alpha=0.1$. This would absorb the propagation of spin waves, reducing the spin-wave reflection at the sample edges. As seen in Fig. \ref{fig:edgedefects}d, the increased damping lead to a broadening of the FMR modes, and suppress some of the splitting of the main mode. The low frequency edge mode however, remains relatively unaffected. 

The edge magnetization $M_s$  was found to be the most important parameter, and we made simulation models where the outer region of the ellipse had a significantly reduced $M_s$ from $M_s = 1.7 \times 10^6$ A/m to  $M_s = 1 \times 10^6$ A/m. Reducing $M_s$ in the edge region changes the FMR spectrum considerably, as seen when comparing Fig. \ref{fig:edgedefects}a and c. The splitting of the main mode is suppressed, and the resonance frequency of the edge mode is shifted significantly. The resulting spectrum now resembles the experimental data, showing mainly two parallel dispersing modes. 

Another important effect to consider in arrays of nanomagnets is the dipolar interaction among the individual particles. In order to take this into account, we performed simulations for arrays of interacting ellipses.

\subsubsection{Dipolar interactions}

The simulations so far have been performed for single ellipses. However, due to the periodic array of ellipses (as shown in Fig. \ref{fig:geom}a), there will be some degree of dipolar interaction between the individual ellipses. 
The dipolar interaction in arrays of magnetic particles can have both static and dynamic contributions. The effects of static dipolar interaction on the magnetization reversal of the same samples have been investigated previously, and an interaction field in the order of tens of mT was found \cite{dipolar}. 
The dynamic interaction can couple the magnetization dynamics of adjacent dots through the stray field generated by the precessing magnetization, forming collective spin excitations in the system \cite{magnonics_review, magnonics_review2}. 

Interactions were included in the simulations by using periodic boundary conditions (b.c.), with the same periodicity as that indicated in Fig. \ref{fig:geom}a. 
In the limit of strong dipolar interaction, one could also expect collective  modes in the system. A simple model of a single ellipse with periodic b.c.  would not be sufficient to resolve such modes, as the neighboring ellipses could rotate either in phase (acoustic mode) or out of phase (optic mode) \cite{magnonic_modes,magnonic_modes2}. To take this into account, we compared the simulation results for a single ellipse with periodic b.c. versus arrays of $3 \times 3$, $5 \times 5$ and $10 \times 10$ ellipses. Comparing the simulated FMR spectrums for the various array sizes, we found no indication of such collective modes in our system. In the following simulations the dipolar interaction was thus taken into account by using a simple model for a single ellipse with periodic b.c. 

\begin{figure}[H]
\centering
\includegraphics[height=100mm, width=85mm]{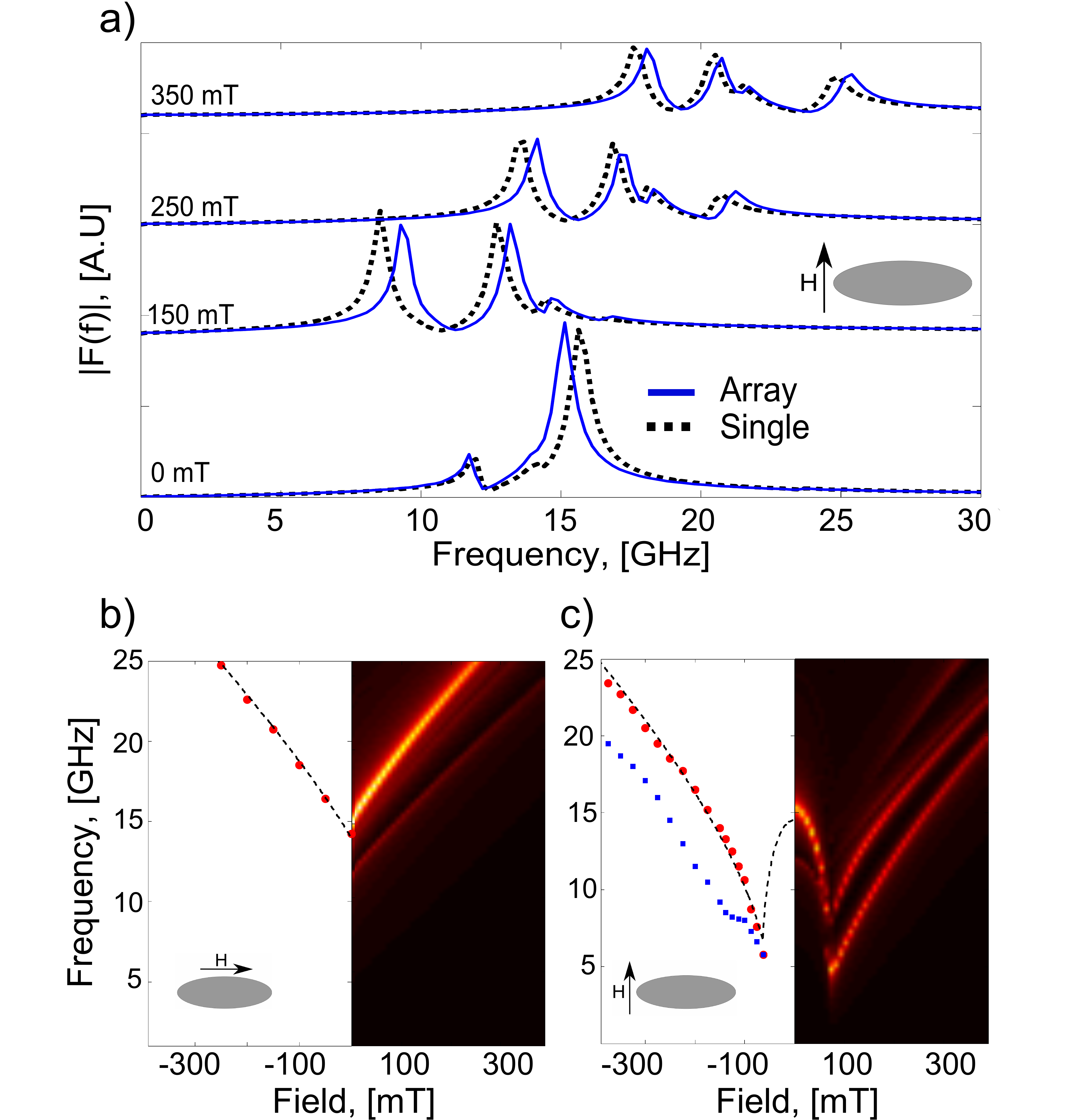} 
\caption{\footnotesize (Color online ) a)  FFT spectrum of  a single ellipse (black dots) vs. an array of ellipses (blue line) using periodic b.c, for an applied field oriented along the short axis of the ellipse.
b)  Left half: Experimental data and macrospin model as dotted black line. Right half: Micromagnetic simulation including edge defects and dipolar interactions in an array of ellipses. Data shown for field oriented along the long axis. c) same data for field oriented along the short axis }
\label{fig:final_model}
\end{figure}

Comparing the simulated spectrums for a single ellipse versus an array of ellipses, we found that the dipolar interaction changes the effective field felt by the individual ellipses. At zero applied field, the magnetization is oriented along the long axis of the ellipses. The overall dipolar field caused by the array geometry will then oppose the magnetization direction. As seen in Fig. \ref{fig:final_model}a, the dipolar field reduces the resonance frequency at zero applied field for the array compared to a single ellipse. Increasing the field along the short axis of the ellipse, the magnetization will reorient itself along the short axis at an applied field of approx. 75 mT (seen as a "dip" in the FMR spectrum in Fig. \ref{fig:edgedefects}c ).
At fields above this switching field, the dipolar interaction acts to increase the effective magnetic field felt by the ellipses, and thus increases the FMR frequency. These shifts can be seen in Fig. \ref{fig:final_model}a for an applied field between 150 mT - 350 mT, and are in the order of 1 GHz. These shifts in the FMR frequencies along the hard/easy axis are similar to those observed by Carlotti \textit{et al.} \cite{interactions}, who studied the effects of dipolar interactions in arrays of rectangular permalloy dots.  

To capture all significant effects we thus made a simulation model with periodic b.c., where edge defects were modelled as a reduced $M_s$ at the sample edges. After including both edge defects and dipolar interactions, one can compare the simulated and experimental spectrums in Fig. \ref{fig:final_model}b and c.

We notice that the inclusion of edge defects and dipolar interactions give a better agreement between the simulated and experimental FMR spectrum. As expected, the edge modes are strongly influenced by edge defects in the samples. 
To accurately capture the behavior of all the FMR modes, it is thus important to take edge defects into account in the simulation model. Due to the large spacing between the individual ellipses in the array, the dipolar interaction is quite weak. In the simulations we observe a small shift in the FMR frequencies, but not any indications of collective modes between neighboring ellipses.  

The fact that the amplitude of the main mode dominates in the experiments, together with the weak dipolar coupling, explains the good agreement between the analytical macrospin model and experimental data. This indicates that in the limit of weak dipolar interaction, our macrospin model can be used to estimate the FMR frequency of the main mode in magnetic elements within the investigated size range (e.g. in a single domain state). Using an analytical macrospin model compared to performing numerical simulations simplifies the analysis considerably. The various energy terms contributing to the FMR dynamics can then be separated, and their relative importance investigated.

\section{Conclusions}
In this study we have investigated how the combined interplay between shape anisotropy and crystalline anisotropy affects the magnetodynamic properties of confined magnetic elements. We have shown how the dimensions of the magnetic elements can be used to balance crystalline and shape anisotropies, and that this can be used to tailor the magnetodynamic properties

We have shown that a simple macrospin model for the FMR frequency gives good agreement with the experimental results for the main FMR mode. 
Comparing experimental data and model calculations, we show how changing the sample size affects the magnetodynamic properties. For the smallest ellipses, shape anisotropy is dominating, whereas for the largest ellipses crystalline anisotropy is the dominating energy term. 
From Eq.(\ref{eq:res4}), the relative contributions to the resonance frequency from crystalline and shape anisotropy is given by: $H_k/\mu_0 M_s (N_x - N_y)$, determined by the anisotropy field $H_k$, the saturation magnetization $M_s$ and the demagnetization factors $N_i$. This means that for the case of a 10 nm thick epitaxial Fe film, one has an intermediate regime between approximately 50 nm to 1 $\mu$m where one can use the sample size as an additional tuning parameter for the dynamic properties. For other materials with a different $H_k$ and $M_s$, this regime can be shifted to smaller/larger sample sizes.

The effects of non ideal sample edges and dipolar interaction in the array of ellipses were investigated using micromagnetic simulations. We found that edge defects in the form of a reduced edge magnetization had to be included in the micromagnetic model, and that this needs to be taken into account in understanding the full FMR spectrum. The static dipolar interaction in the array was found to shift the FMR frequency in the order of 1 GHz compared to that of a single ellipse. From the simulated FMR spectrums we found no indications of collective spin excitations due to the dynamic dipolar interaction between neighboring ellipses. 

The tunability of the relative contributions from crystalline and shape anisotropies means that by changing the material parameters and sample size one can tailor the magnetodynamic properties of the magnetic elements, which could be of importance for magnonics applications.

\section*{Acknowledgements}
This work was supported by the Norwegian Research Council (NFR), project number 216700. 
V.F acknowledge partial funding obtained from the Norwegian PhD Network on Nanotechnology for Microsystems, which is sponsored by the Research Council of Norway, Division for Science, under contract no. 221860/F40.
 F.M.  acknowledges  support from Catalan Government COFUND-FP7. JMH and FM also thank  the Spanish  Government  (Grant  No.   MAT2011-23698.)

\end{document}